\documentclass[final,5p,times,twocolumn]{elsarticle}

\usepackage{amssymb}

\journal{Astroparticle Physics}

\begin{document}

\begin{frontmatter}



\title{Ultrahigh Energy Nuclei in the Turbulent Galactic Magnetic Field}


\author[label1,label2]{G.~Giacinti}
\ead{giacinti@apc.univ-paris7.fr}
\author[label3]{M.~Kachelrie\ss}
\author[label1,label4]{D.~V.~Semikoz}
\author[label2]{G.~Sigl}

\address[label1]{AstroParticle and Cosmology (APC), Paris, France}
\address[label2]{II. Institut f\"ur Theoretische Physik, Universit\"at Hamburg, Germany}
\address[label3]{Institutt for fysikk, NTNU, Trondheim, Norway}
\address[label4]{Institute for Nuclear Research of the Russian Academy of Sciences, Moscow, Russia}

\begin{abstract}
In this work we study how the turbulent component of the Galactic magnetic
field (GMF) affects the propagation of ultrahigh energy heavy nuclei. We
investigate first how the images of individual sources and of the 
supergalactic plane depend on the properties of the turbulent GMF. Then we 
present a quantitative study
of the impact of the turbulent field on (de-) magnification of source fluxes,
due to magnetic lensing effects. We also show that it is impossible to explain
the Pierre Auger data assuming that all ultrahigh energy nuclei are coming from
Cen~A, even in the most favorable case of a strong, extended turbulent field
in the Galactic halo.
\end{abstract}


\end{frontmatter}


\section{Introduction}

The composition of the cosmic ray (CR) flux above energies $E\gtrsim 10^{17}$\,eV
is a topic of current debate. The shape of the CR
spectrum~\cite{hires-spec,Abraham:2008ru,Abraham:2010mj} at the highest energies is compatible with either a proton or a heavy nuclei composition. On the one hand, the correlation of ultrahigh energy cosmic rays (UHECRs) arrival
directions with the large scale distribution of active galactic nuclei reported by the Pierre Auger 
collaboration~\cite{auger-anisotropy,correlation:2010zzj} could be
consistent with a composition dominated by protons. On the
other hand, the possibility to accelerate nuclei to higher energies as
protons may favour a transition to heavier nuclei at the end of the UHECR spectrum.
Meanwhile, various results of measurements of the UHECR composition have been presented: The Pierre Auger Observatory air shower measurements
suggest a gradual shift towards a heavier composition starting at a few times
$10^{18}$\,eV up to $\simeq4\times10^{19}\,$eV above which the
statistics is currently still insufficient for composition
studies~\cite{Collaboration:2010yv}.
The Yakutsk EAS Array muon data are compatible with this 
result~\cite{Glushkov:2007gd}. However, both HiRes 
measurements~\cite{BelzICRC} and preliminary studies of the Telescope
Array~\cite{TA} do not agree with this shift. The results of their composition
studies favour proton primaries.

In this work, we extend our previous analysis of the implications of a possible heavy primary
composition on the UHECRs propagation in the Galactic magnetic
field (GMF)~\cite{Giacinti:2010dk}: We add a {\em turbulent\/} component to the regular GMF 
and investigate in details its influence. We present in this paper simulations for iron
nuclei with the energy $E= 60$\,EeV, since this is the highest energy
where present experiments collect still a reasonable number of events.

T.~Stanev suggested one of the first models of the regular GMF including a
spiral arm-like structure of the field in the Galactic
disk~\cite{Stanev:1996qj}. Other models were later proposed in Refs.~\cite{Harari:1999it,Tinyakov:2001ir}. 
The authors of Refs.~\cite{Vallee} and~\cite{Page:2006hz} presented a toroidal 
field consisting of concentric rings and another axisymmetric field, 
respectively. Spiral patterns based either on
the structure of the NE2001 thermal electron density model~\cite{Brown:2007qv}
or on the spiral structure of the Milky Way~\cite{Jiang:2010yc} deduced from HII
regions and giant molecular clouds~\cite{Hou:2009gn} have also been proposed to
describe this field. The implications on the GMF modelling of recent rotation
measure maps were reported in~\cite{Han:2006ci,Han:2009jg,Han:2009ts}. Some of
the most recent constraints on the regular disk field have been presented
in~\cite{Rae:2010jy,VanEck:2010jz,VanEck:2010ka}.

The authors of Ref.~\cite{PS} presented first a GMF model adding a halo
contribution made of toroidal and poloidal fields to the disk
field~\cite{PS,Kachelriess:2005qm}. In the present paper, we refer to this model as 
the ``PS model''. Several new GMF models were proposed and
confronted with the data in Refs.~\cite{Sun:2007mx,Sun:2010sm,Pshirkov:2011um}. At
present, no theoretical GMF model can fit all experimental
data~\cite{Jansson:2009ip,Waelkens:2008gp}.



Earlier works discussing the propagation of UHECRs in turbulent fields include
Refs.~\cite{Harari:2002dy,Tinyakov:2004pw}. The first studied the
generic theoretical properties of multiple image formation and
amplification in a turbulent magnetic field for point-like UHECR
sources~\cite{Harari:2002dy}. Ref.~\cite{Tinyakov:2004pw} discussed
the range of values one could expect for the relative ratios of UHECR
deflections in the regular and turbulent GMF. In our present work, we
deal with the case of iron nuclei at the highest energies emitted by
extended extragalactic sources considering both the regular and the
turbulent components of the GMF.

Our paper is structured as follows: Section~\ref{MethodsField}
presents and discusses the two numerical methods that we use to
generate the turbulent GMF. In
Section~\ref{ConsequencesSourcesearches}, we investigate qualitatively
the consequences induced by the turbulent GMF on the shape of images
of single sources and of the supergalactic plane. The implications of
this field on the (de-) magnification of the fluxes from single sources are
presented in Section~\ref{MagnificationBR}. In
Section~\ref{OneSource}, we rule out the possibility that all presently
observed UHECR events could have been emitted by only one nearby
source and deflected over the celestial sphere by a strong turbulent-dominated GMF.

\section{Modelling of the turbulent Galactic magnetic field}
\label{MethodsField}

In this section, we present the two methods we use to model the
turbulent Galactic magnetic field and recall some of its basic properties.
A turbulent magnetic field $\textbf{B}$ satisfies $\left\langle\textbf{B}(\textbf{r})\right\rangle=\textbf{0}$ and
$\left\langle\textbf{B}^2(\textbf{r})\right\rangle \equiv B_{\rm rms}^2>0$. At the energy we consider throughout this work, $E=60$\,EeV, the Larmor radius of iron primaries is typically much larger 
than the coherence length $L_c$ of the turbulent field. As a result,
the effect of a uniform turbulent field can be parametrised in this
regime by $L_c$ and $B_{\rm rms}$. Therefore, our results for the deflections 
are largely independent of the exact shape of the fluctuation spectrum  
which is poorly constrained. In the following we will consider for 
definiteness a Kolmogorov spectrum. If $k$ denotes the modulus
of wave vectors, the power spectrum of such a field is $\mathcal{P}(k)\propto
k^{-5/3}$ and the amplitudes $\textbf{B}(\textbf{k})$ of its Fourier modes
follow $|\textbf{B}(\textbf{k})|^2\propto k^{-11/3}$.

The modulus of wave vectors $\textbf{k}$ in Fourier space satisfy
$2\pi/L_{\max}\le k=|\textbf{k}|\le2\pi/L_{\min}$, where $L_{\min}$ and $L_{\max}$ 
denote the minimal and the maximal scales of variation present in the field 
$\textbf{B}(\textbf{r})$, respectively. The correlation length  $L_c$ of the field is equal to 
\begin{equation}
L_c = \frac{1}{5} \: L_{\max}\: \frac{1-(L_{\min}/L_{\max})^{5/3}}{1-(L_{\min}/L_{\max})^{2/3}}~.
\end{equation}

Turbulent magnetic fields can be generated either directly in coordinate space
as a superposition of a finite number of plane waves, or by computing the 
field on a three-dimensional grid using the Fast Fourier Transform (FFT). 
In the present paper we use the first method in Sections~\ref{ConsequencesSourcesearches}
and~\ref{MagnificationBR} following Ref.~\cite{JG99}. 
The field $\textbf{B}(\textbf{r})$ is written as the sum of $N_m$ modes,
\begin{equation}
\textbf{B}(\textbf{r}) = \sum_{n=1}^{N_m} A_n \varepsilon_n \exp \left( i{\bf
    k}_n\cdot{\bf r}+ i \beta_n \right)~,
\label{EqnBfield}
\end{equation}
where $\varepsilon_n = \cos(\alpha_n)
\widehat{\textbf{x}}_n + i \sin(\alpha_n) \widehat{\textbf{y}}_n$,
$\alpha_n$ and $\beta_n$ represent random phases and the unit vectors
$\widehat{\textbf{x}}_n$ and $\widehat{\textbf{y}}_n$ form an
orthogonal basis together with the randomly chosen direction of the
n-th wave vector $\widehat{\textbf{k}}_n$. The amplitude $A_n$ of the n-th plane wave is given by
\begin{equation}
A_n^2=\frac{B_{\rm rms}^2G(\textbf{k})}{\sum_{n=1}^{N_m} G(\textbf{k}_n)}~,
\end{equation}
with
\begin{equation}
G(\textbf{k})=\frac{4 \pi k^2 \Delta k}{1+(k L_c)^{11/3}}~.
\end{equation}

We choose $k_n$ between $k_{\min}=2\pi/L_{\max}$ and $k_{\max}=2\pi/L_{\min}$ 
with equal spacings in logarithmic scale: $\Delta \log k = \Delta k /
k={\rm const.}$ The number of modes is set to $N_m=1000$. We verified
that with this we obtain the theoretically expected deflections 
in the small-angle scattering regime with an error smaller than $\simeq2\%$. 
Once the modes are dialed, the turbulent field can be computed with
Eq.~(\ref{EqnBfield}) at any point in space with arbitrary precision
for the given realization. Its drawback is that 
computations are much slower than with the second method described below, 
because the method requires to compute many trigonometric functions at each 
step of the particle trajectory.

The second method consists in precomputing the field on a three dimensional grid
in real space. The coordinates of its wave vectors correspond to positions
of vertices on the reciprocal grid in the Fourier space. The resulting field in
real space is computed with the Fast Fourier Transform (FFT)~\cite{NR}. The
magnetic field can be quickly interpolated at any point of the space. We take an
8~point-interpolation of the field values on the eight vertices of the grid cell
in which the considered point lies. 

We use this method in Section~\ref{OneSource} to derive results which require
the propagation of $3.6\times10^6$ particles. For this work, using a three dimensional cubic
grid of $N=2^8$ vertices per side is sufficient. It contains $256^3$ values
for each component of the turbulent field. This grid of $\simeq2.6$\,kpc lateral
size is periodically repeated in real space. 

In the three following sections, we use the same type of generic profile for the rms
strength of the turbulent field, $B_{\rm rms}(r,z)$. We take the model introduced in
Ref.~\cite{Giacinti:2009fy},
\begin{equation}
B_{\rm rms}(r,z) = B(r) \exp\left(-\frac{|z|}{z_{0}}\right)~,
\end{equation}
where $r$ is the galactocentric radius and $z$ the distance to the
Galactic plane. The parameter $z_{0}$ characterizes
the extension of the random field in the halo. In the next sections, we will
vary
$z_{0}$ in the range 0.75\,kpc to 10\,kpc. The radial profile $B(r)$ is defined as 
\begin{equation}
B(r) = \left\{ \begin{array}{ll}
                B_{0} \: \exp{\left( \frac{5.5}{8.5} \right) }  & \mbox{, if } r \leq 3\,\mbox{kpc (bulge)}\\
		B_{0} \: \exp{\left( \frac{ - \left( r - 8.5\,{\rm kpc}\right)}{8.5\,{\rm kpc}} \right) } & \mbox{, if } r > 3\,\mbox{kpc}
               \end{array} \right.
\end{equation}
where $B_{0}$ denotes the value of $B_{\rm rms}$ close to the Sun.
For the regular field, we use the PS model choosing the parameters 
as described in Ref.~\cite{Giacinti:2010dk}.

\section{Qualitative consequences of deflections in the turbulent GMF for source searches}
\label{ConsequencesSourcesearches}

\begin{figure*}[!t]
  \centerline{\includegraphics[width=0.46\textwidth]{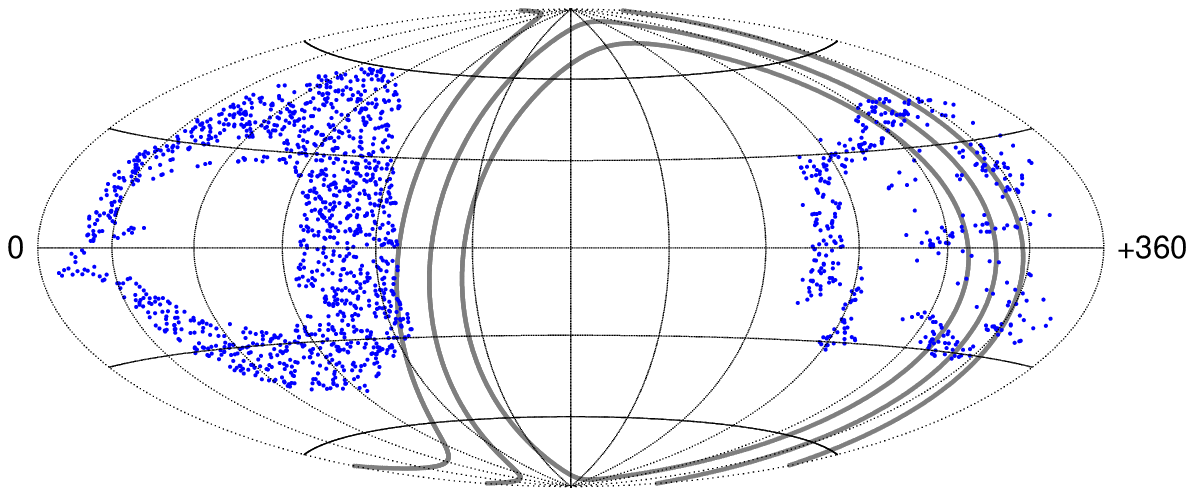}
              \hfil
              \includegraphics[width=0.46\textwidth]{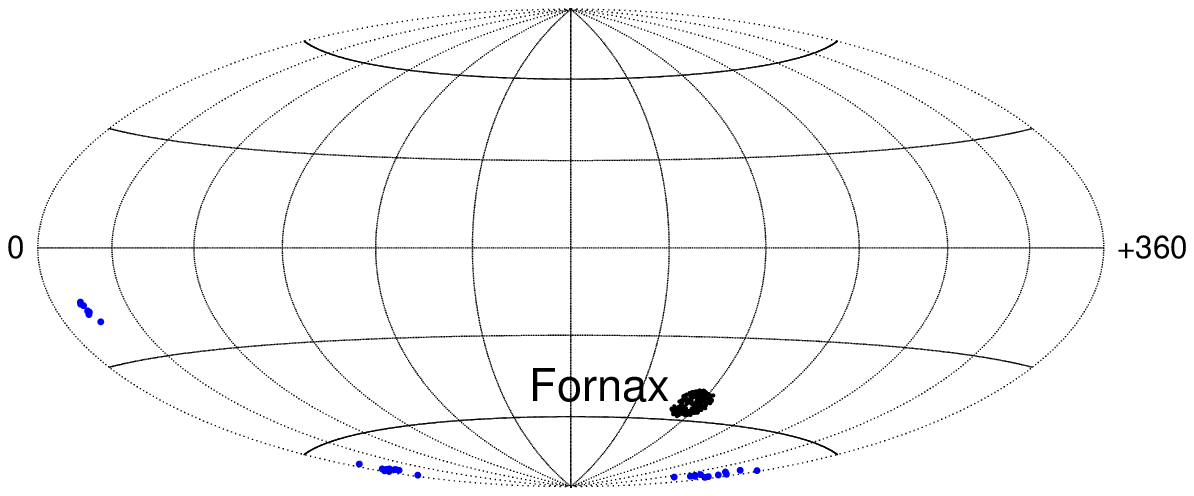}
             }
  \caption{Images of the supergalactic plane \textbf{(left panel)} and
  of the Fornax galaxy cluster \textbf{(right panel)}, respectively
  considered as a $\pm10^{\circ}$ band in supergalactic latitude and a
  5$^{\circ}$ radius disk, emitting 60\,EeV iron nuclei deflected in
  the PS regular GMF model. No turbulent magnetic field was
  added. Dark blue points represent arrival directions of cosmic
  rays. The grey lines stand for the supergalactic plane and the
  $\pm10^{\circ}$ circles in supergalactic latitude. The black disk
  represents Fornax. The maps are in Galactic coordinates with the anti-Galactic
  center in the center. Increasing $l$ from the left to the right.}
  \label{RegGMF}
\end{figure*}

We investigate here qualitatively the influence of the turbulent GMF 
on the images of single sources and the supergalactic plane emitting
60\,EeV iron nuclei. Since the characteristic properties
of the turbulent GMF are still poorly known, we test the dependence 
of the results on its main parameters.

The shape of proton and light nuclei source images has been discussed in
Refs.~\cite{Golup:2009cv,Giacinti:2009fy}. We have shown in
Ref.~\cite{Giacinti:2010dk} how images of extragalactic UHECR sources would look
like at Earth for different models of the regular GMF for a heavy primary
composition. Contrary to proton sources, one expects that some
iron sources have several images even at the highest energies. Moreover,
some of them may be strongly magnified or
demagnified~\cite{Harari:1999it,Harari:2000az,Harari:2000he,Giacinti:2010dk}. We
investigate below the impact of a non-zero turbulent GMF contribution on such
results.

We use the model discussed in Section~\ref{MethodsField} for the profile of the
rms strength of the turbulent field $B_{\rm rms}$. For the regular GMF, we use the
PS model as a generic example. Images of given UHECR sources can strongly vary
from one GMF model to another. However, global properties of the whole sky such
as mean deflection angles or fractions of the sky with given
(de-)magnification~\cite{Giacinti:2010dk} are similar in most recent regular GMF
models. In this respect, the PS model is representative for a generic regular 
GMF model.

Large scale structure MHD simulations agree that extragalactic magnetic fields tend to
be strongest in the large scale structure, around the largest matter
concentrations~\cite{Dolag:2003ra,Dolag:2004kp,Sigl:2004yk,Das:2008vb,Ryu:2009pf}.
However, they disagree on some aspects such as the filling factor distributions
(\textit{i.e.} the fraction of space filled with fields above a given strength,
as a function of that strength~\cite{Sigl:2004gi}). This leads to substantial
differences on the sizes of UHECR deflections predicted by different models,
ranging from negligible~\cite{Dolag:2003ra,Dolag:2004kp} to more than ten
degrees for proton primaries, even at the very highest energies~\cite{Sigl:2004yk,Sigl:2004gi}. Despite
this, even in the case of strong deflections, UHECR would still
preferentially arrive from directions towards the large scale structure
(LSS) of the galaxy distribution. Since the fields in the voids are very small, UHECR arrival
directions outside the Galaxy should point back to the supergalactic plane, even
if they do not point back to their sources. Therefore, if sources are located in
the LSS and if extragalactic deflections are sizable, it is sufficient in a
first approximation to study the image of the supergalactic plane itself.

\begin{figure*}[!t]
  \centerline{\includegraphics[width=0.46\textwidth]{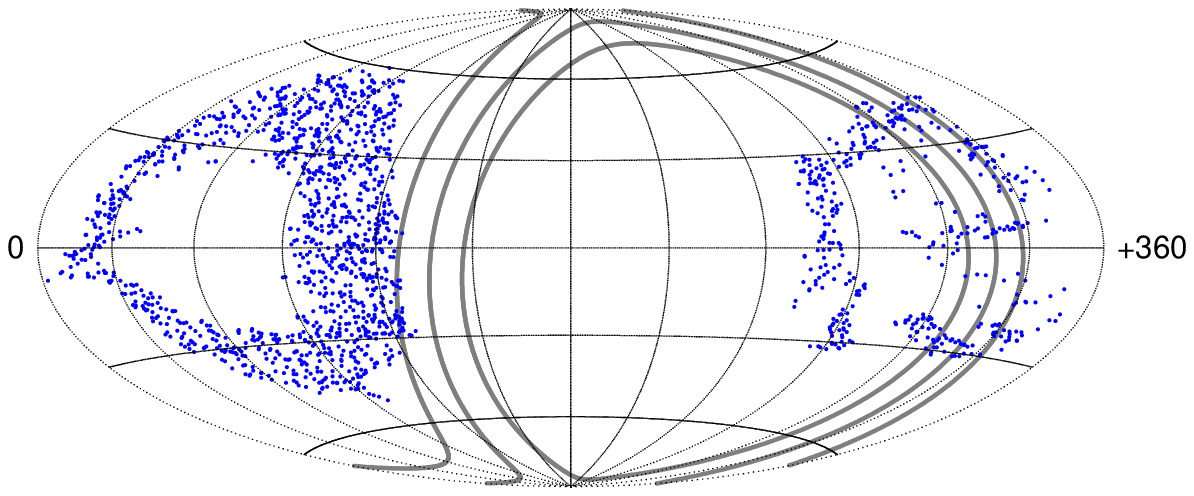}
              \hfil
              \includegraphics[width=0.46\textwidth]{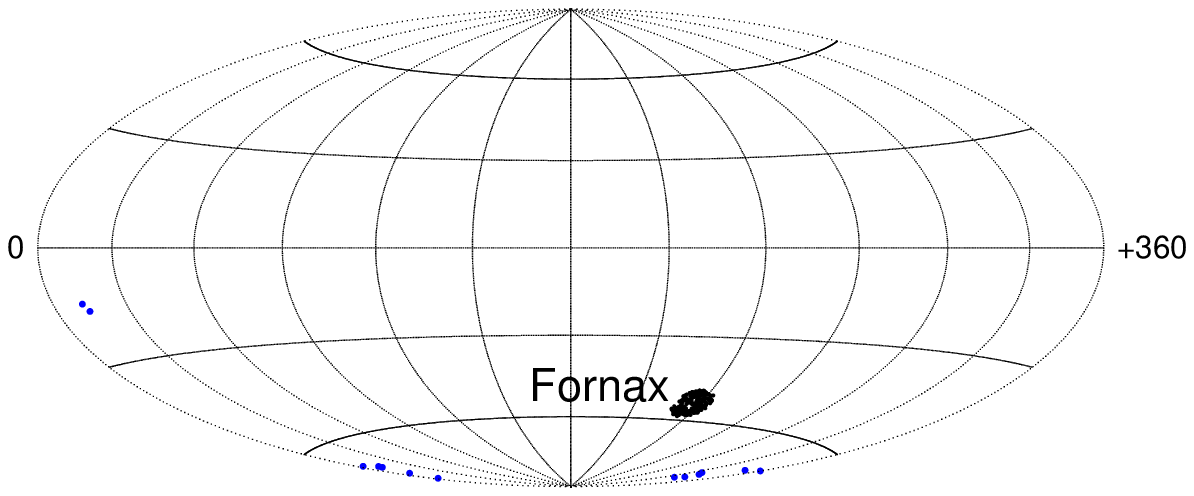}
             }
  \centerline{\includegraphics[width=0.46\textwidth]{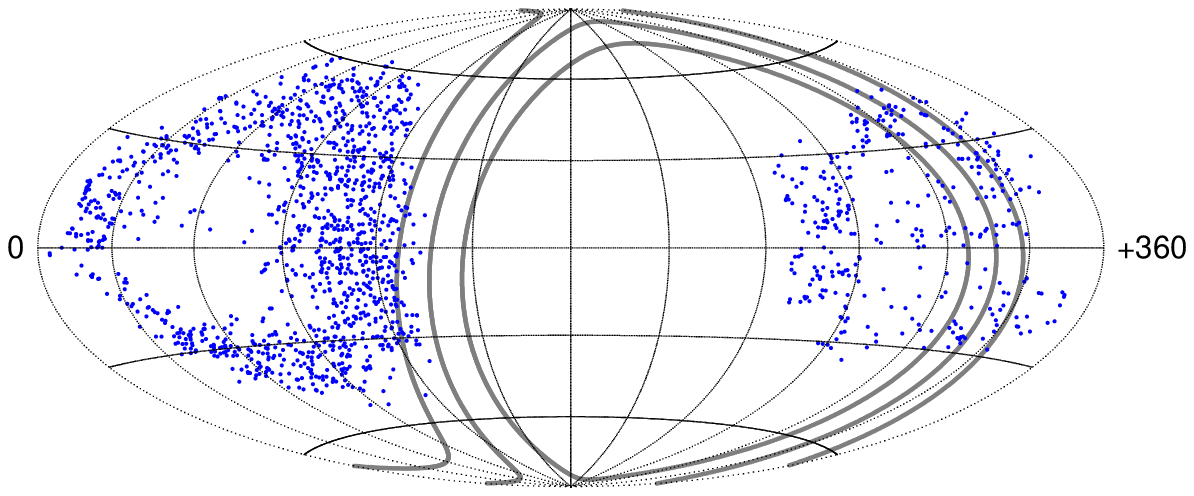}
              \hfil
              \includegraphics[width=0.46\textwidth]{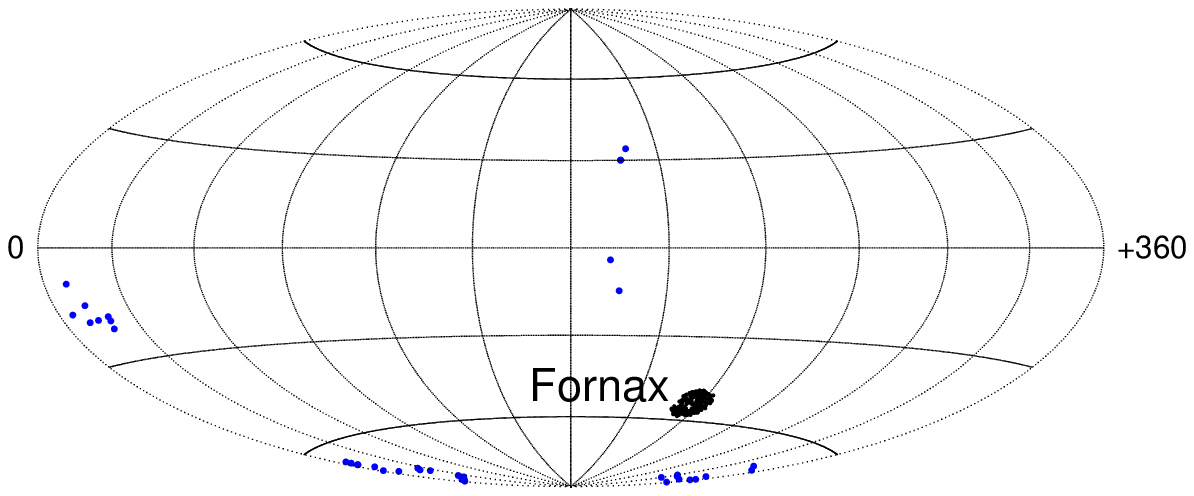}
             }
  \centerline{\includegraphics[width=0.46\textwidth]{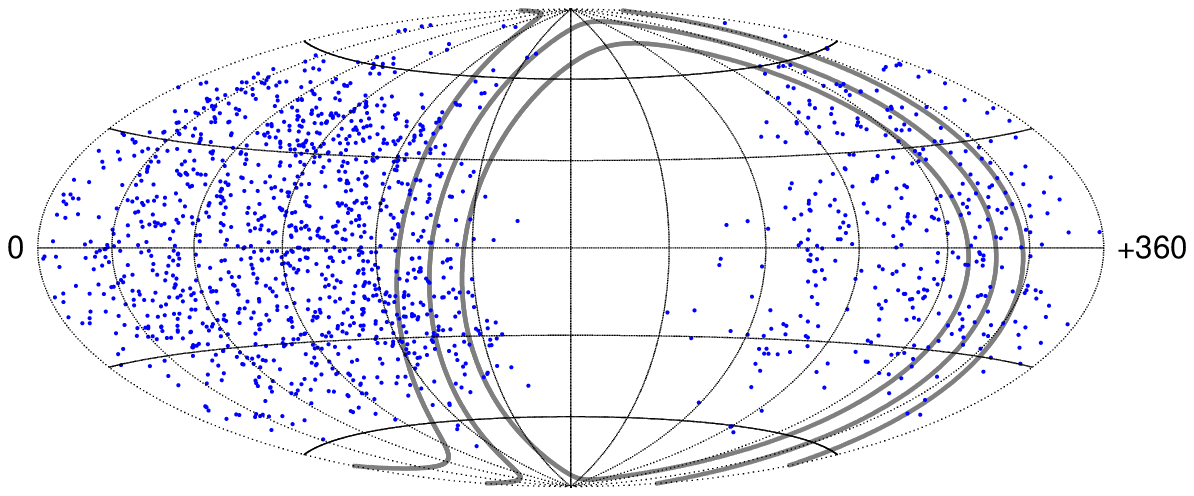}
              \hfil
              \includegraphics[width=0.46\textwidth]{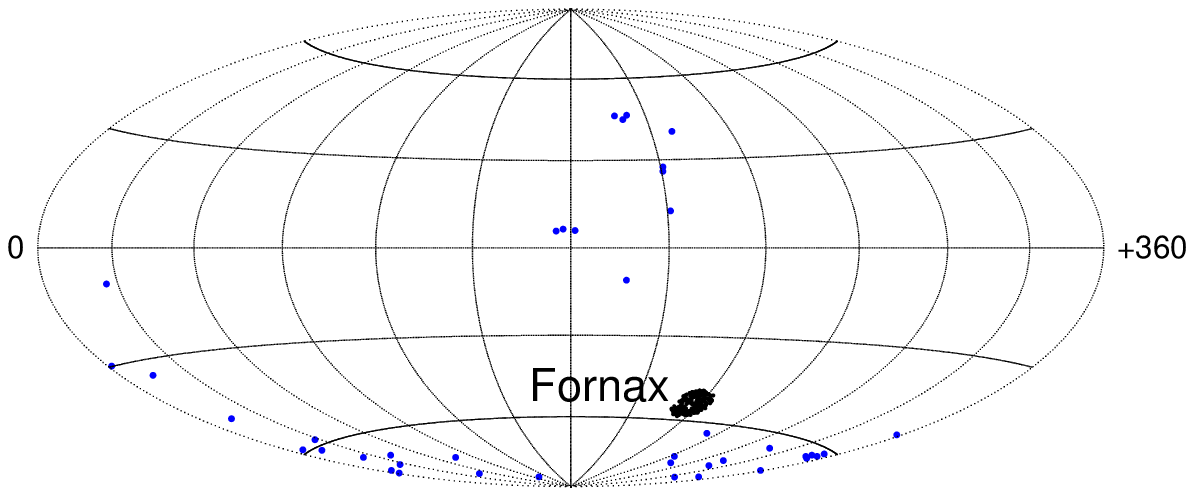}
             }
  \caption{Impact of the turbulent field strength at Earth $B_0$ on the
  60\,EeV iron nuclei images of the supergalactic plane \textbf{(left
    column)}, and of the Fornax galaxy cluster \textbf{(right
    column)}. Same GMF regular component as for Fig.~\ref{RegGMF} (PS
  model), with in addition a non-zero turbulent Galactic magnetic
  field. It has a Kolmogorov spectrum of correlation length
  $L_{c}=50$\,pc. Its extension into the halo is $z_{0}=3$\,kpc. See text for details on
  the field profile. \textbf{First row:} $B_0=0.25\,\mu$G;
  \textbf{Second row:} $B_0=1\,\mu$G; \textbf{Third row:}
  $B_0=4\,\mu$G. Same key as for Fig.~\ref{RegGMF}.}
  \label{Brms}
\end{figure*}

\begin{figure*}[!t]
  \centerline{\includegraphics[width=0.46\textwidth]{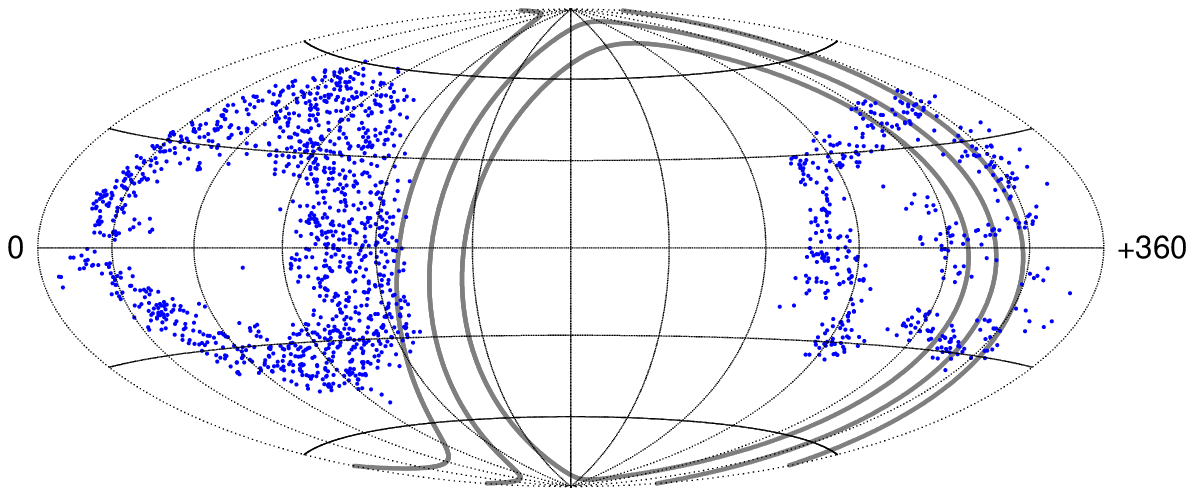}
              \hfil
              \includegraphics[width=0.46\textwidth]{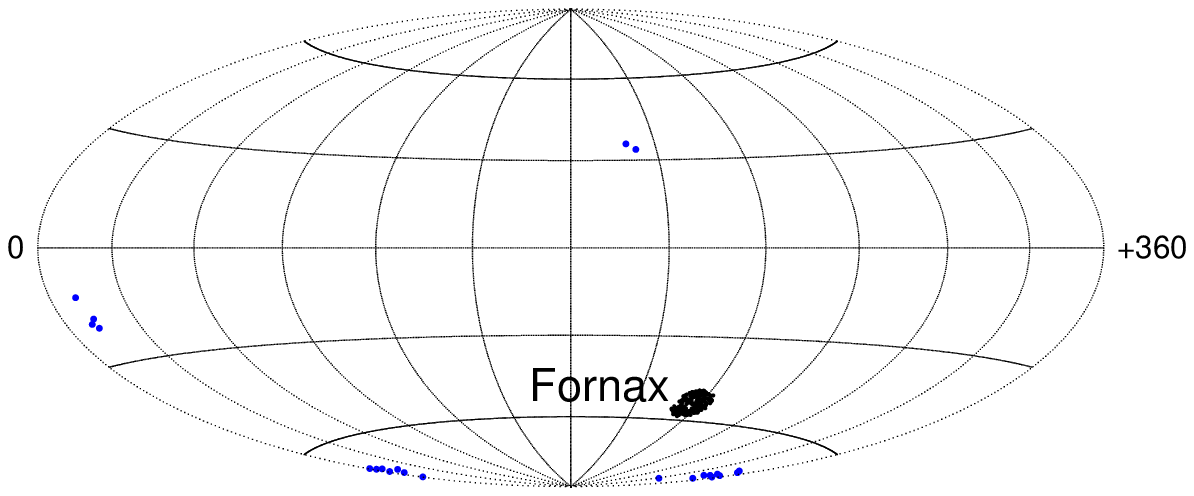}
             }
  \centerline{\includegraphics[width=0.46\textwidth]{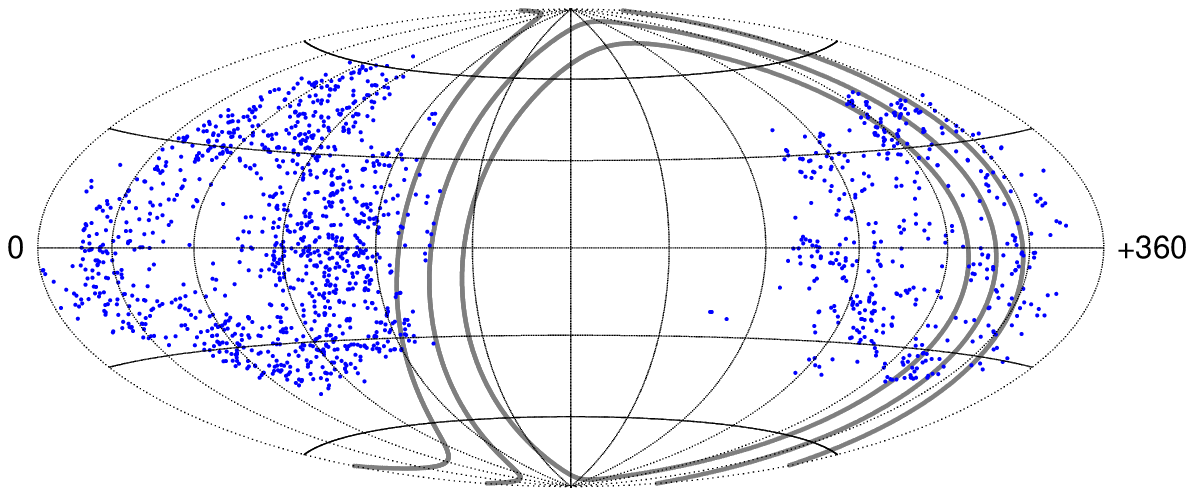}
              \hfil
              \includegraphics[width=0.46\textwidth]{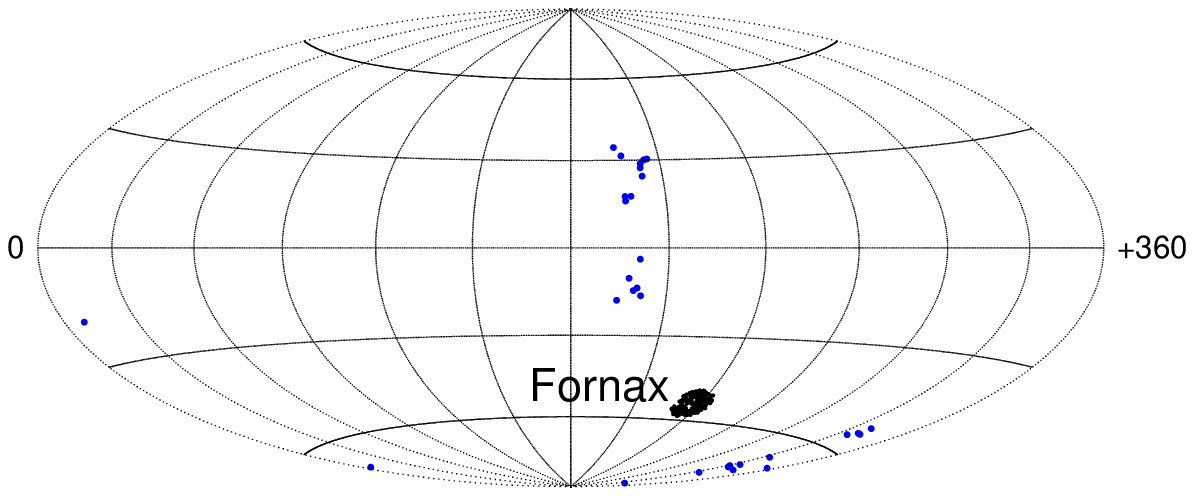}
             }
  \caption{Columns and key as in Fig.~\ref{Brms}. Impact of the
  turbulent Galactic magnetic field correlation length $L_{c}$ on the
  supergalactic plane and Fornax 60\,EeV iron images. Turbulent field
  with a Kolmogorov spectrum, $B_0=1\,\mu$G and $z_{0}=3$\,kpc. See
  text for details on the field profile. \textbf{First row:}
  $L_{c}=12.5$\,pc;
  \textbf{Second row:} $L_{c}=200$\,pc. PS model for the regular Galactic magnetic
  field.}
  \label{Lc}
\end{figure*}

\begin{figure*}[!t]
   \centerline{\includegraphics[width=0.46\textwidth]{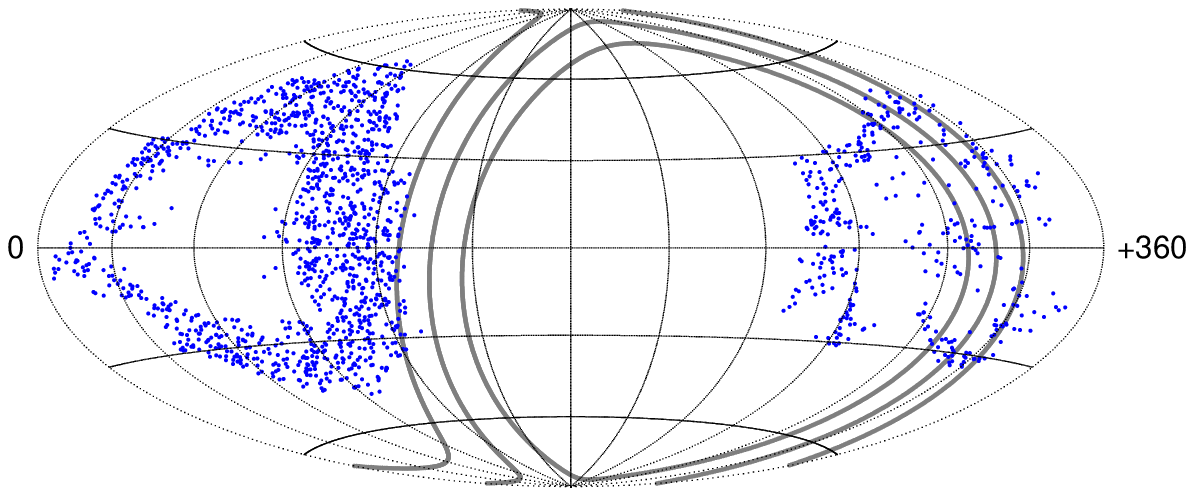}
              \hfil
              \includegraphics[width=0.46\textwidth]{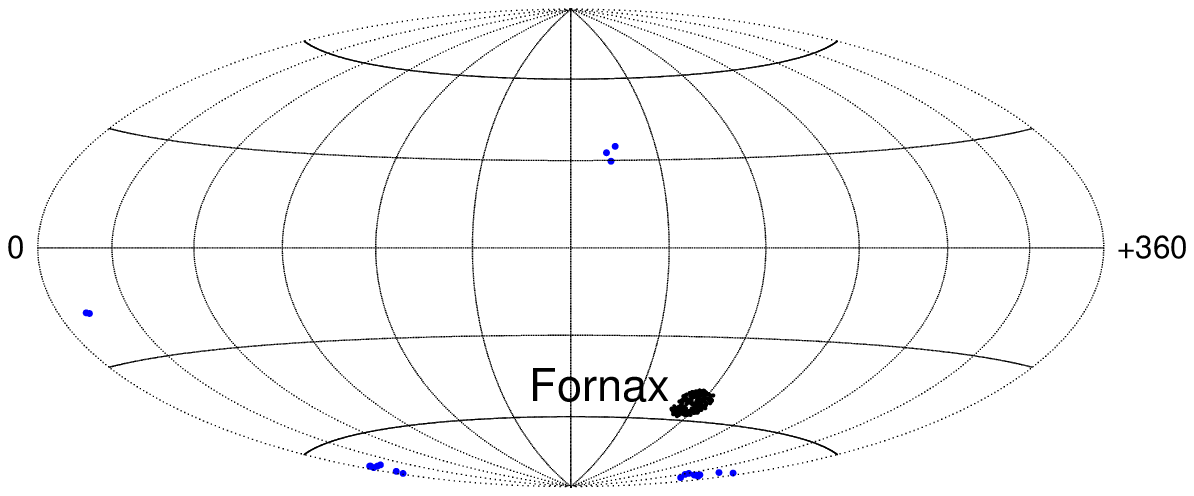}
             }
   \centerline{\includegraphics[width=0.46\textwidth]{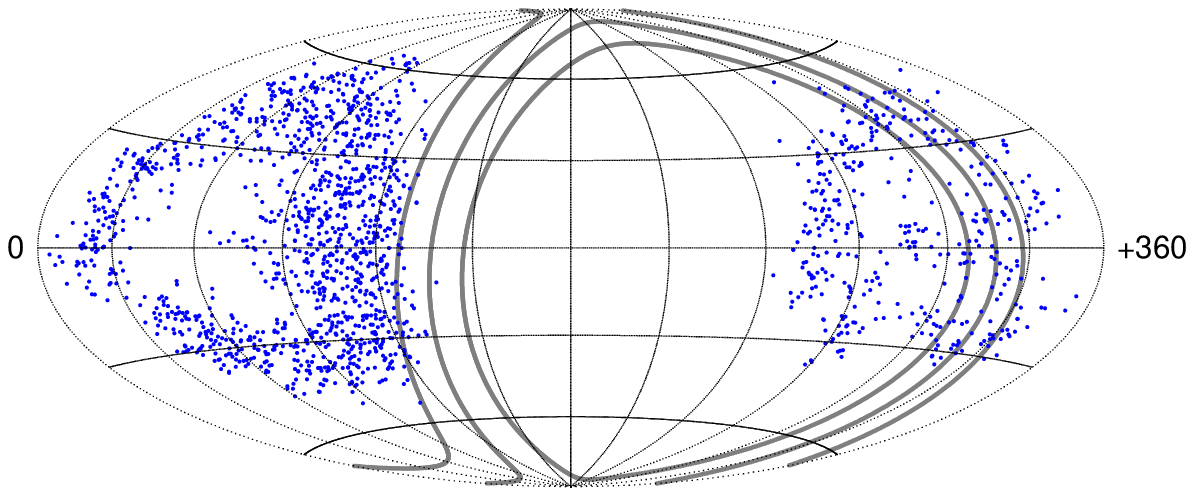}
              \hfil
              \includegraphics[width=0.46\textwidth]{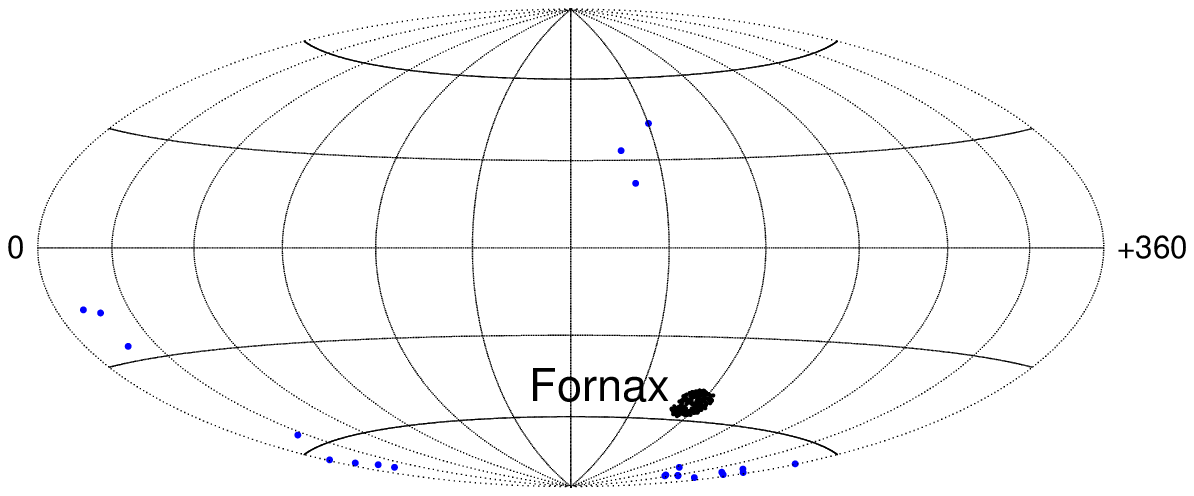}
             }
  \caption{Columns and key as in Fig.~\ref{Brms}. Impact of the
  turbulent Galactic magnetic field extension into the halo $z_{0}$ on
  the supergalactic plane and Fornax 60\,EeV iron images. Turbulent
  field with a Kolmogorov spectrum of correlation length
  $L_{c}=50$\,pc, and
  $B_0=1\,\mu$G. See text for details on the field
  profile. \textbf{First row:} $z_{0}=0.75$\,kpc; \textbf{Second row:}
  $z_{0}=6$\,kpc. PS model for the regular Galactic magnetic field.}
  \label{Z0}
\end{figure*}

We show in Figs.~\ref{RegGMF}--\ref{Z0} the 60\,EeV iron
nuclei images of the whole supergalactic plane (left columns) and of a specific
galaxy cluster, Fornax (right columns).

For the images of the supergalactic plane, we assume 
that the region containing most of the local distribution of matter has
supergalactic latitudes $b_{SG}$ between $\pm 10$ degrees. Cosmic ray arrival
directions at Earth are represented in the figures by dark blue points, and the
three grey lines correspond to the supergalactic plane ($b_{SG}=0^{\circ}$) and
the $b_{SG}=\pm 10^{\circ}$ circles. The 60\,EeV iron image of
  the supergalactic plane in the PS model is split into two. The
  regions of the supergalactic plane near both Galactic poles cross
  blind regions and therefore do not contribute to the image observed
  at Earth, see Fig.~\ref{Amplification} (upper left panel). This cuts
  the supergalactic plane (SGP) into two parts: The part of the plane located at $l\sim120^{\circ}-150^{\circ}$ mostly produces the image at $l\sim0^{\circ}-130^{\circ}$, and the part at $l\sim300^{\circ}-330^{\circ}$ is responsible for the image at $l\sim250^{\circ}-350^{\circ}$. At energies $E\gtrsim120$\,EeV, the two images start to merge. In the recent GMF models presented in Refs.~\cite{Sun:2007mx,Pshirkov:2011um}, the 60\,EeV iron nuclei images of the SGP display several common features. The largest part of the image(s) also appear in the region $l\sim0^{\circ}-180^{\circ}$, as for the PS model, and the largest contributor to this image is the part of the SGP which is located at $l\sim120^{\circ}-150^{\circ}$. In all tested models, the overall geometry of the image in the $l\sim0^{\circ}-180^{\circ}$ region is ``circle-like'' as for the PS model. The image may be enlarged and shifted on the sky by more than $30^{\circ}$ from one model to another, and its details are model-dependent.

For the galaxy cluster images, we choose the Fornax cluster because its high
Galactic latitude (in absolute values) avoids complicated images due to the
magnetic field in the Galactic disk. In the PS model, 60\,EeV iron images from
the two nearby clusters with higher latitudes, Virgo and Coma, turn out to be
invisible at Earth due to their strong demagnification~\cite{Giacinti:2010dk}.
Fornax is regarded here as a source extending over a 5$^{\circ}$ radius. Even in the
case the cluster would contain only one source, deflections by magnetic 
fields inside the cluster lead to an extended source~\cite{Dolag:2008py}. 
For the galaxy cluster images, we also assume that deflections of
UHECR nuclei in extragalactic magnetic
fields are weak enough (as in, e.g.\ the simulations of 
Refs.~\cite{Dolag:2003ra,Dolag:2004kp}) that the UHECRs are not spread 
out over large regions of the LSS. In the figures, the cluster is denoted by a
black disk.

To generate these figures, we backtrace $10^4$ iron antinuclei emitted
isotropically from the Earth to the outside of the Galactic halo. We
record those which escape the Galaxy in
directions in the $b_{SG}=-10^{\circ} \mbox{~to~} +10^{\circ}$ band for the
supergalactic plane images, and within 5 degrees from the center of Fornax for
the Fornax images.

Figure~\ref{RegGMF} presents the supergalactic plane and Fornax images,
deflected in the PS regular GMF only. This figure can be considered as the
reference to which Figs.~\ref{Brms}--\ref{Z0} should be compared.

For Figs.~\ref{Brms},~\ref{Lc} and~\ref{Z0}, a non-zero turbulent GMF component
was added to the PS regular component. They show the dependence of the 
deflections on the turbulent field rms amplitude $ B_0$, its correlation length $L_{c}$ and
its extension above the Galactic plane $z_{0}$. For clarity, we choose a
given ``reference'' configuration of the turbulent field parameters
$\left\lbrace B_0,L_{c},z_{0}\right\rbrace$, and for each figure, only one
parameter varies, while the two others are fixed to the reference
values for which we take $B_0=1\,\mu$G, $z_{0}=3$\,kpc and
$L_{c}=50$\,pc. This configuration corresponds to the second row 
in Fig.~\ref{Brms}. For all examples presented in this work, we adopt a Kolmogorov spectrum with $L_{\min}=0.4\,L_c$ and $L_{\max}=4\,L_c$.

Each plot shown in this section is computed for one given realisation
of the turbulent field. We have checked that the results would not be qualitatively different if one takes another realisation of the turbulent GMF. The only exception to this
is the lower row in Fig.~\ref{Lc}, for the turbulent field with $L_{c}=200\,$pc
and $z_{0}=3$\,kpc. Since $z_{0}$ is not much larger than $L_{\max}$, cosmic 
rays are more sensitive to the given realisation of the field.

In Fig.~\ref{Brms}, each row corresponds to a different strength of the
turbulent GMF. From the top to the bottom, the field value at Earth
$B_0$ is respectively set to $0.25\,\mu$G, $1\,\mu$G and $4\,\mu$G. The
effect of this parameter is easily visible both in the Fornax and supergalactic
plane images. When the turbulent field strength is increased, the images are
more and more spread out, compared to those in Fig~\ref{RegGMF}. For example, the
angular spread of the Fornax image at ($l\sim0^{\circ}$,$b\sim-70^{\circ}$) 
increases
from negligible to nearly comparable to the magnitude of the deflection in the
regular field itself. At the same time, the ``center'' of the image of Fornax
stays at the same place as in the regular field. On the two last rows, a few
cosmic rays are even deflected to the northern Galactic hemisphere. For the case
$B_0=4\,\mu$G, the structure of the supergalactic plane image becomes
practically unrecognizable. For this regular GMF model and the
parameters of the turbulent component considered here, deflections are
however still not sufficiently large to spread the image over the
whole celestial sphere.

Fig.~\ref{Lc} presents the impact of the correlation length $L_{c}$. In the
first row, $L_{c}$ is 4 times smaller than the reference value,
$L_{c}=12.5$\,pc. Images are less extended
and more compact than in the case $L_{c}=50$\,pc. They nearly resemble to the
images in Fig.~\ref{RegGMF}, computed without any turbulent field. On the second
row, $L_{c}$ is set to 200\,pc. The spread of the images is larger.

In Fig.~\ref{Z0}, we investigate the impact of the turbulent field extension into
the Galactic halo. The first and second rows respectively represent the cases
$z_{0}=0.75$\,kpc and $z_{0}=6$\,kpc. The fine structure of the images is indeed
more spread for $z_{0}=6$\,kpc than for $z_{0}=0.75$\,kpc, but the effect is
less remarkable than the dependence on $B_0$ in Fig.~\ref{Brms}. 

The mean deflections in the turbulent field grow with $B_0$,
$L_{c}$ and $z_{0}$. In the limit case of propagation in the ballistic 
regime, the rms deflection $\delta_{\rm rms}$ for particles of rigidities $E/Z$ 
propagating the distance $L$ in a turbulent field of constant rms strength $B_{\rm rms}$ is equal to 
\begin{displaymath}
\delta_{\rm rms} = \frac{1}{\sqrt{2}} \frac{ZeB_{\rm rms}}{E}\left(LL_{c}\right)^{1/2}
\end{displaymath}
\begin{equation}
\simeq 6.3^{\circ} \times \frac{Z}{26} \frac{60\,\mbox{EeV}}{E} \frac{B_{\rm rms}}{1\,\mu\mbox{G}}\left(\frac{L}{3\,\mbox{kpc}}\right)^{1/2}\left(\frac{L_{c}}{50\,\mbox{pc}}\right)^{1/2}~.
\label{eqTC}
\end{equation}

If both the source and its image are located on the same side of the Galactic plane and at high enough Galactic latitudes, the distortion of the image follows the simple expectation of Eq.~(\ref{eqTC}). The image of Fornax at ($l\sim0^{\circ}$,$b\sim-70^{\circ}$) is a good illustration of this dependence on the turbulent field parameters -see Figs.~\ref{RegGMF}--\ref{Z0}. In this case, iron nuclei from one given source with $E\gtrsim60$\,EeV are more or less spread around the arrival directions that they would have had in the regular field only. Because of the linear dependence on $B_{\rm rms}$, the size of the spread is more sensitive to a relative change of $B_0$ than to a change of $L_{c}$ or $z_{0}$.

Contrary to this simple case, if the source or its image is located at ``low'' Galactic latitudes ($|b|\lesssim30-40^{\circ}$), even relatively small changes of trajectory due to the turbulent field can lead to very large distortions of the image. Some images may even appear in completely different locations. Such a behavior can be seen for the image of Fornax at ($l\sim15^{\circ}$,$b\sim-15^{\circ}$), see Fig.~\ref{RegGMF}. Depending on the parameters of the turbulent magnetic field, this image is partly -or completely- shifted in other parts of the sky, and new images with Galactic latitudes $|b|\lesssim40^{\circ}$ appear -see Figs. \ref{Brms}--\ref{Z0}. Such effects cannot be described by the simple analytical formula of Eq.~(\ref{eqTC}). A full study requires the numerical computations conducted in this work.

For the ranges of turbulent GMF parameters considered here, we find no
noticeable qualitative difference on source images between a field with a
Kolmogorov spectrum and a field with the same modulus for all wave vectors $\textbf{k}$
in Fourier space, \textit{i.e.} $L_{\min}=L_{\max}=2L_{c}$, as long as the
fields have the same correlation length.

For such parameters, iron nuclei from a source with $E\gtrsim60$\,EeV are not sufficiently spread to cover the
whole sky. However, if one takes the extreme case of a very
strong field, $B_0=10\,\mu$G, extending very far into the halo,
$z_{0}=10\,$kpc, both the images of single sources and of the supergalactic
plane are spread over the whole sky. Therefore, we investigate in
Section~\ref{OneSource} if the Pierre Auger data published recently~\cite{correlation:2010zzj}
may be compatible with only one nearby UHECR source, such as Cen~A, in case 
of a very strong and extended turbulent GMF.

\section{Flux (de-) magnification of sources and extension of blind regions}
\label{MagnificationBR}

 \begin{figure*}[!t]
   \centerline{\includegraphics[width=0.46\textwidth]{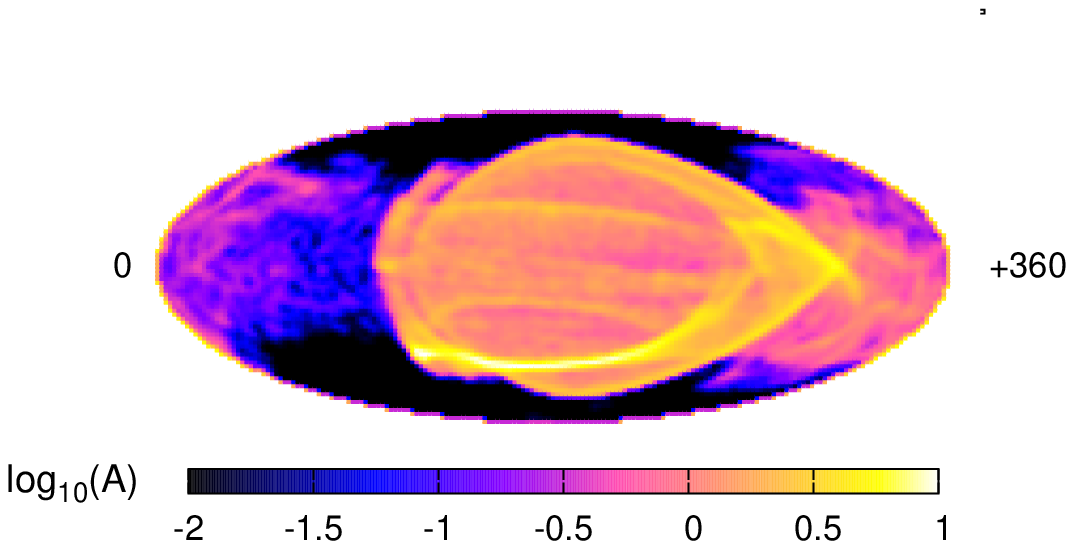}
              \hfil
              \includegraphics[width=0.46\textwidth]{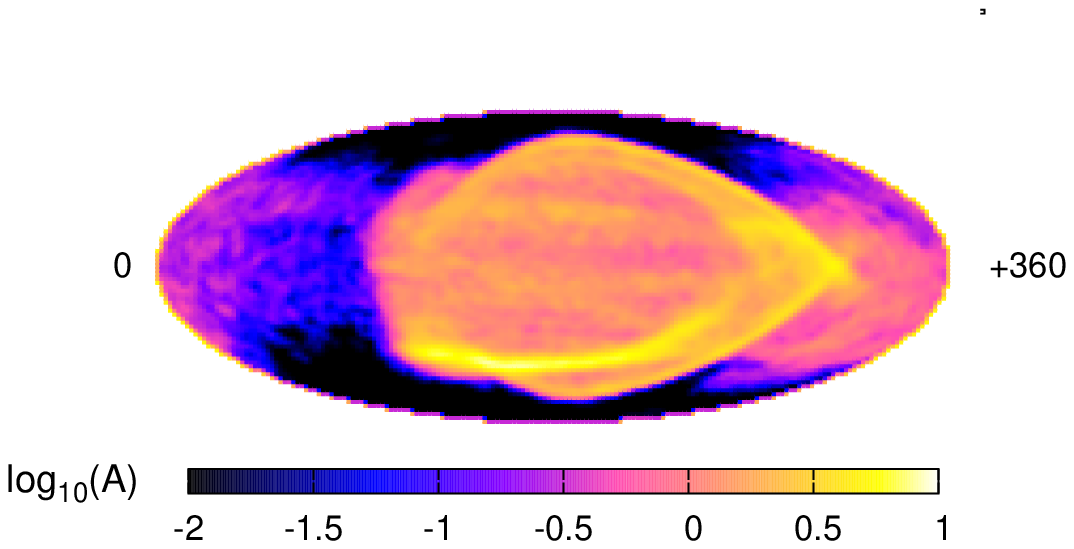}
             }
   \centerline{\includegraphics[width=0.46\textwidth]{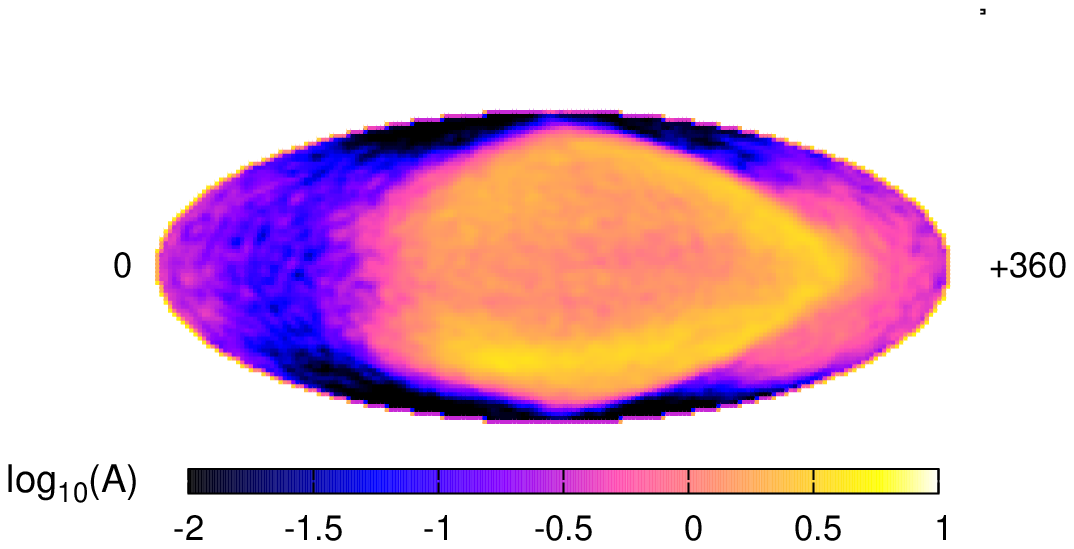}
              \hfil
              \includegraphics[width=0.46\textwidth]{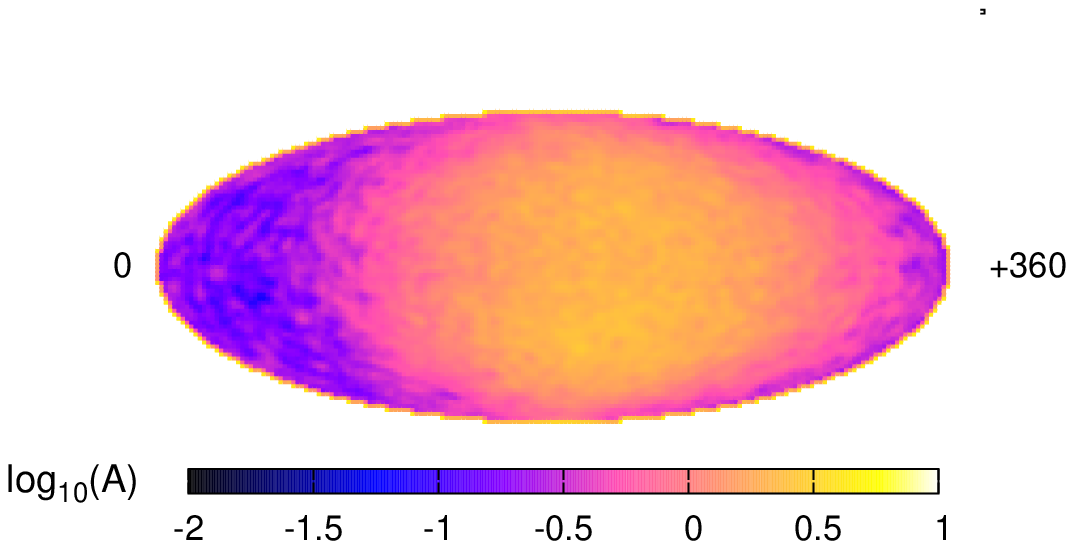}
             }
   \caption{Logarithm of the amplification factor $\mathcal{A}$ of the
  flux of extragalactic 60\,EeV iron nuclei sources, depending on
  their position on the sky. Plots are in Galactic coordinates, with
  the Galactic anti-center in the center. We take the PS model for the
  regular component of the GMF. Regions with
  $\log_{10}(\mathcal{A})\leq-2$ are in black. \textbf{Upper left
    panel:} No additional turbulent field component is added. The
  small fraction of the sky with $\log_{10}(\mathcal{A})\geq1$ is in
  white; \textbf{Upper right panel:} Additional turbulent field with
  $B_0=1\,\mu$G, $L_{c}=50$\,pc and $z_{0}=3$\,kpc; \textbf{Lower left
    panel:} $B_0=4\,\mu$G, $L_{c}=50$\,pc and $z_{0}=3$\,kpc;
  \textbf{Lower right panel:} $B_0=10\,\mu$G, $L_{c}=50$\,pc and
  $z_{0}=10$\,kpc.}
   \label{Amplification}
 \end{figure*}

 \begin{figure*}[!t]
   \centerline{\includegraphics[width=0.33\textwidth]{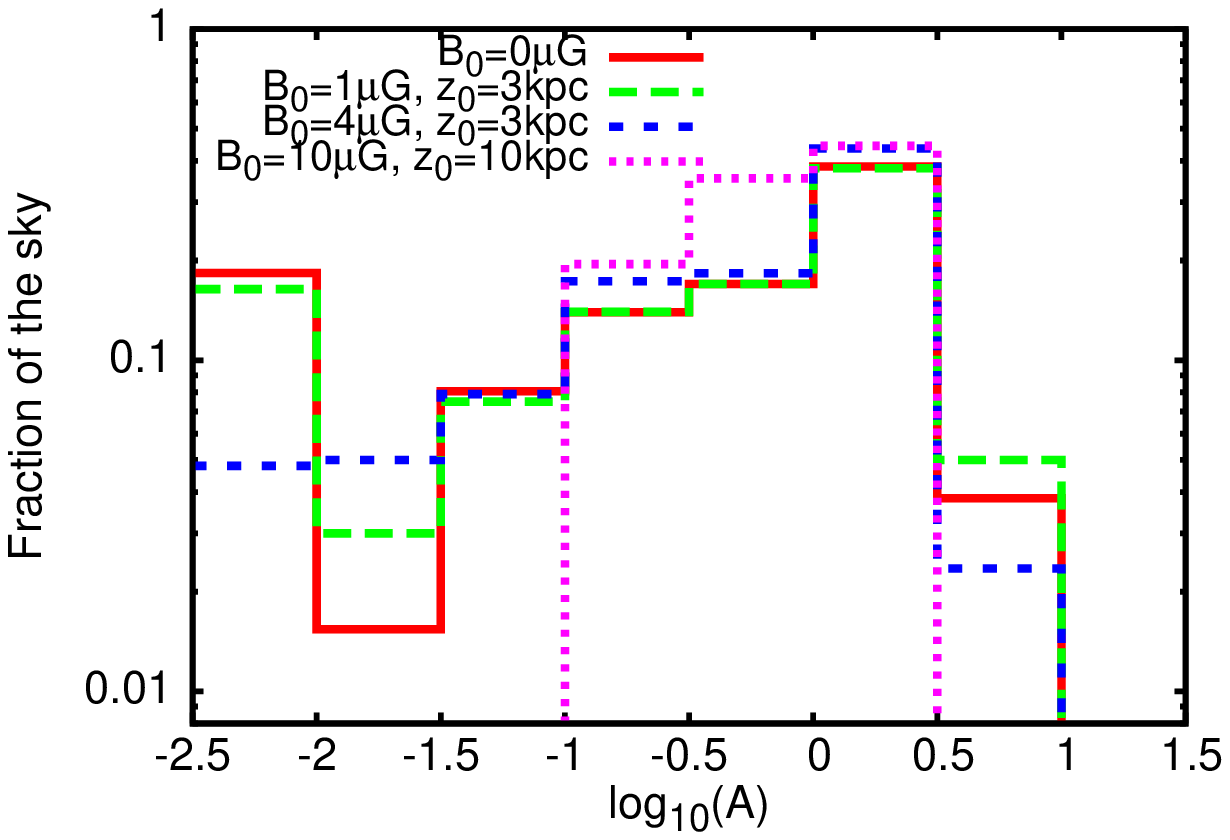}
              \hfil
              \includegraphics[width=0.33\textwidth]{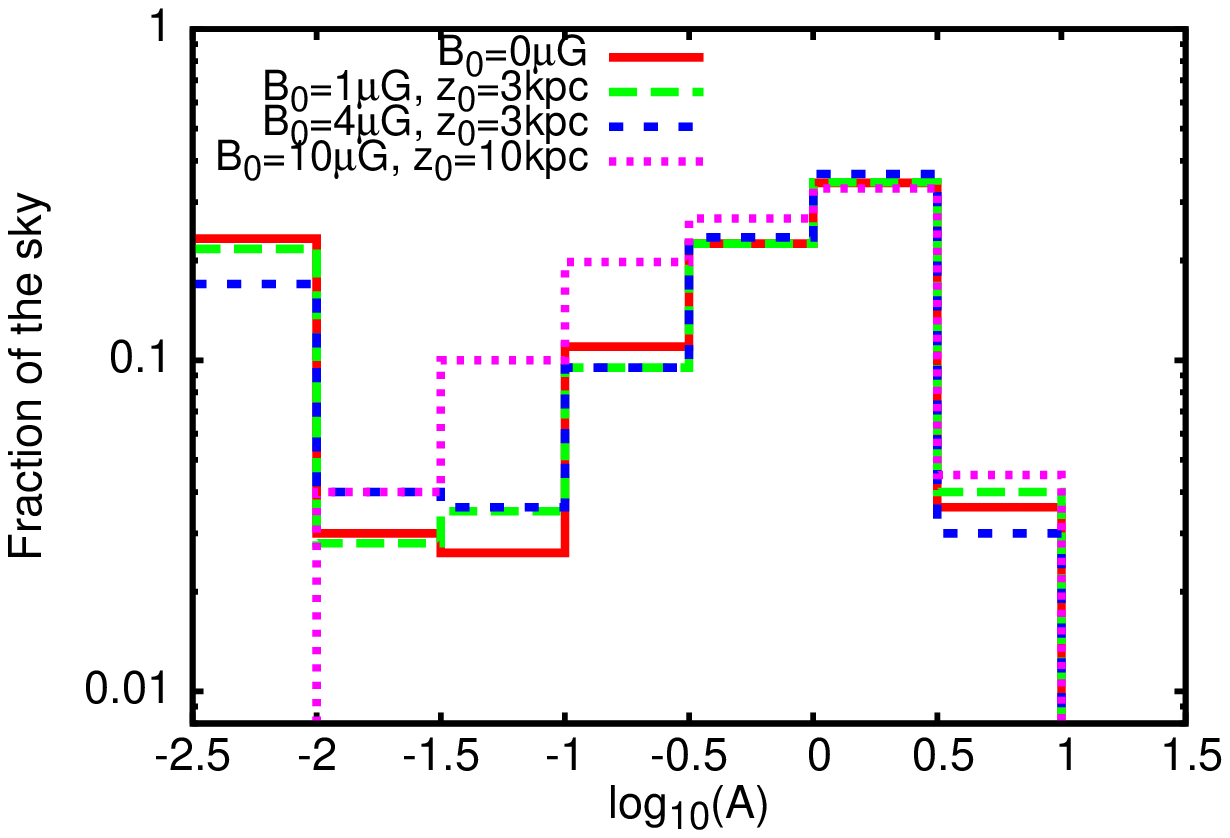}
              \hfil
              \includegraphics[width=0.33\textwidth]{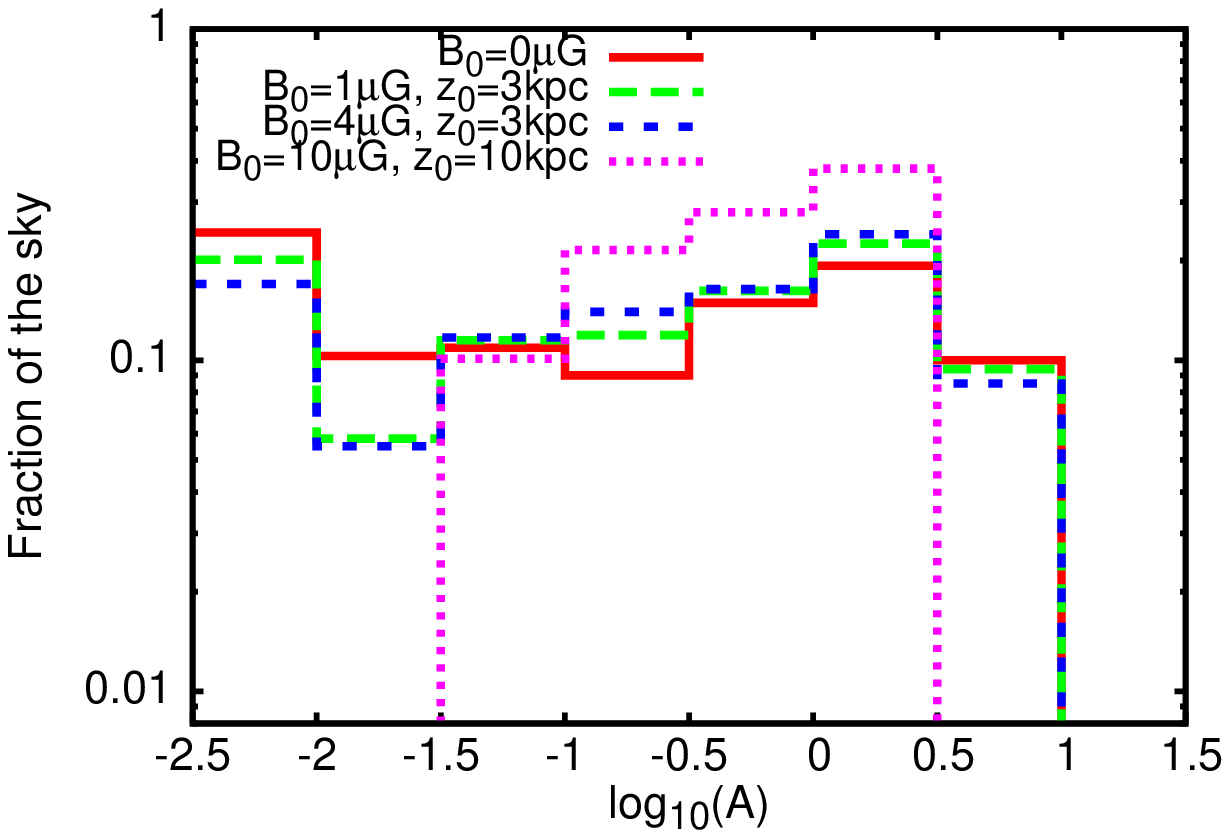}
             }
   \caption{Histogram of fractions of the sky outside the Galaxy with
  amplifications $\mathcal{A}$ within bins of width
  $\Delta\log_{10}(\mathcal{A})=0.5$. \textbf{Left panel:} For all four histograms, the GMF
  regular component is given by the PS model; \textbf{Middle panel:} Sun08 model -``ASS+RING'' model of Ref.~\cite{Sun:2007mx}; \textbf{Right panel:} Sun08-MH model, as defined in Ref.~\cite{Giacinti:2010dk}. In each plot, solid red lines for no additional turbulent field component. Green, blue and magenta lines for the
  cases of a turbulent field with parameters respectively set to
  ($B_0=1\,\mu$G, $z_{0}=3$\,kpc), ($B_0=4\,\mu$G, $z_{0}=3$\,kpc) and
  ($B_0=10\,\mu$G, $z_{0}=10$\,kpc). The correlation length L$_{c}$ is
  50\,pc. The bin of amplifications $\mathcal{A}$ below $10^{-2}$ corresponds to blind regions.}
   \label{Diagram}
 \end{figure*}

In this section, we investigate the impact of the turbulent GMF on the (de-)
magnification of individual source fluxes. Magnification and demagnification
effects are due to magnetic lensing of UHECR in the GMF. In
Ref.~\cite{Giacinti:2010dk}, we studied the contribution of the regular GMF
component. We have shown that in the case of iron primaries, source fluxes can
be strongly magnified or demagnified even for energies as high as 60\,EeV.
In particular, we found large regions of the sky which can be considered 
empty for a given number of backtraced nuclei: For instance, regions with an 
under-density $\rho/\langle\rho\rangle < 10^{-2.5}$ can encompass 
one fifth of the sky. Thus sources located in these parts of the sky would 
not be detectable at Earth by current and next generation UHECR experiments. 
We now study how these previous conclusions are affected by the turbulent 
component of the GMF. 

The sky maps in Fig.~\ref{Amplification} show the logarithm of the flux
amplification factor $\mathcal{A}$ of extragalactic 60\,EeV iron nuclei sources,
depending on their positions on the sky. We take the PS model for the regular GMF. We use the same method as in Ref.~\cite{Giacinti:2010dk} to compute $\mathcal{A}$: We isotropically inject
10$^5$ iron antinuclei at Earth and backtrace them to the border of the Galaxy. Their
relative densities on the celestial sphere outside the Galaxy correspond to
$\mathcal{A}$. These densities are computed on 3$^{\circ}$ radius circles
covering the whole sky.

Let us call here ``blind regions'' the regions of the extragalactic sky in which
$\log_{10}(\mathcal{A})\leq-2$ for a given energy. At this energy, sources
located in such regions of the sky cannot be seen at Earth by current or next
generation UHECR experiments, due to the strong demagnification of their
fluxes. In general, the places, extensions and shapes of blind regions do not change abruptly with energy but vary continuously with energy. For instance the blind regions in the PS model at 80\,EeV are similar to those at 60\,EeV, though slightly less extended. In Fig.~\ref{Amplification}, we plot blind regions in black. The
amplification is below $10^{-2}$ in some places.

The upper left panel in Fig.~\ref{Amplification} represents the
reference plot in which the turbulent field is set to zero. In the
three other panels, a turbulent component with $L_{c}=50$\,pc is added to the
regular field. Its strength and extension in the halo are, respectively, equal to
($B_0=1\,\mu$G, $z_{0}=3$\,kpc), ($B_0=4\,\mu$G, $z_{0}=3$\,kpc) and
($B_0=10\,\mu$G, $z_{0}=10$\,kpc), for the upper right, lower left and
lower right panels. 

Figure~\ref{Amplification}
shows that when the amplitude and the spatial extension of the turbulent GMF 
halo are increased, the extension of regions of extreme magnification and
demagnification tend to shrink. For example, the reduction of the size of the
darkest and brightest regions from the upper left to the lower left panels is
clearly visible. For the lower right panel (extremely strong and extended
turbulent field), regions with $-0.5\leq\log_{10}(\mathcal{A})\leq0.5$ encompass
$\simeq80\%$ of the sky. This means that in $\simeq80\%$ of the sky, 60\,EeV
iron nuclei sources would have their fluxes neither magnified or demagnified by
more than a factor $\simeq3$. In the case without any turbulent field, this
fraction of the sky with moderate flux modification falls to
$\simeq55\%$. Blind regions even disappear in the lower right panel.

Figure~\ref{Diagram} (left panel) shows four histograms corresponding to the four
cases shown in Fig.~\ref{Amplification}. The histograms represent the
fractions of the sky with given amplifications $\mathcal{A}$.
These histograms are computed for 10$^5$ backtraced antinuclei with
the method used for Fig.~8 of Ref.~\cite{Giacinti:2010dk}. $10^5$
antinuclei are sufficient to probe the bin $\log_{10}(\mathcal{A})\leq-2$.
This figure confirms the effect of the turbulent Galactic magnetic
field. As previously pointed out, the distributions tend to peak
around the mean value $\mathcal{A}\sim1$ and have lower variances for
stronger and more extended turbulent components.

In addition, we can compute the maximum and minimum logarithmic
amplifications $\log_{10}(\mathcal{A})$ on the whole sky.
For the case without the turbulent field, we have
$\log_{10}(\mathcal{A})\simeq1.15$ for the maximum amplification. When
including turbulent fields with the parameters
($B_0=1\,\mu$G, $z_{0}=3$\,kpc), ($B_0=4\,\mu$G, $z_{0}=3$\,kpc)
and ($B_0=10\,\mu$G, $z_{0}=10$\,kpc), we respectively find that the
maximum values for $\log_{10}(\mathcal{A})$ are $\simeq0.95$,
$\simeq0.63$ and $\simeq0.47$.
For the case with $B_0=10\,\mu$G and $z_{0}=10$\,kpc, the minimum value
for $\log_{10}(\mathcal{A})$ is $\sim-1.6$. For the other cases, the
minimum amplifications are too low to be probed with only $10^5$
backtraced antinuclei.

The larger the amplitude and extension of the turbulent GMF, the
weaker the amplitude between the maximum magnification and
demagnification. This result is in line with our previous findings.

The fraction of the sky covered by blind regions corresponds to the
left-most bin of the diagram in Fig.~\ref{Diagram} (left panel). Without taking
into account the turbulent field, we find that blind regions encompass
$\simeq18\%$ of the sky for the PS model. When including turbulent
fields with the parameters ($B_0=1\,\mu$G,
$z_{0}=3$\,kpc), ($B_0=4\,\mu$G, $z_{0}=3$\,kpc), ($B_0=10\,\mu$G,
$z_{0}=10$\,kpc), blind regions respectively represent $\simeq16\%$,
$\simeq5\%$ and 0\% of the sky. Therefore, if the turbulent component
is strong and sufficiently extended 
in the $z-$direction, the fraction of blind regions on the sky can easily
fall below 15\%. However, this would not facilitate source searches
due to the bigger spread and blurring of their images, cf. the
previous section.

\begin{figure*}[!t]
   \centerline{\includegraphics[width=0.46\textwidth]{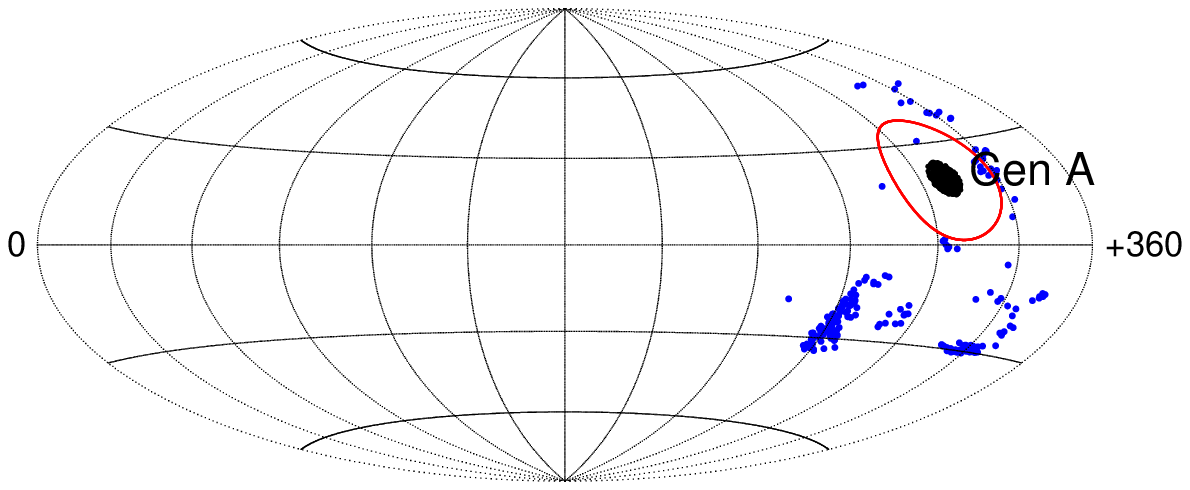}
              \hfil
              \includegraphics[width=0.46\textwidth]{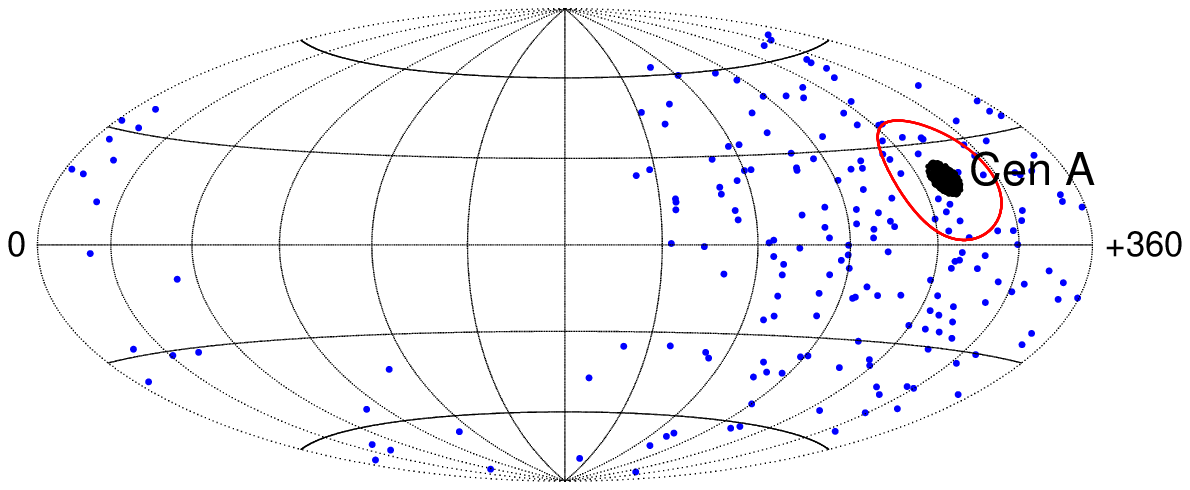}
             }
  \caption{\textbf{Left panel:} Image of iron nuclei with $E\geq55$\,EeV emitted by
  Cen~A, deflected in the PS regular GMF model only, and convolved with the Pierre Auger Observatory exposure;
  \textbf{Right panel:} Same as in the left panel, with an additional strong and extended turbulent
  Galactic magnetic field. The turbulent magnetic field has a Kolmogorov spectrum, with $L_{c}=50$\,pc, $B_0=10\,\mu$G
  and $z_{0}=10$\,kpc. Sky maps in Galactic coordinates, with the
  anti-Galactic center in the center. Cosmic rays are in dark blue and
  the black disks denote the position of Cen~A on the sky. The red circles surround the region within
  $18^{\circ}$ from Cen~A.}
  \label{CenA_Reg_TF}
\end{figure*}

We test to which extent these results are generic. We show in the two other panels of Fig.~\ref{Diagram} the same diagrams for two other regular GMF models. In Fig.~\ref{Diagram} (middle panel), we take the ``ASS+RING'' model of Reference~\cite{Sun:2007mx}, which we will call the ``Sun08'' model. In Fig.~\ref{Diagram} (right panel), we use the modified version of the Sun08 model introduced in Ref.~\cite{Giacinti:2010dk}. We call it the ``Sun08-MH (Modified Halo)'' model. The four diagrams correspond to the same turbulent field strengths and extensions in the halo, as in the left panel. One can see that the sizes of blind regions without turbulent field are very similar in the three models, though they decrease faster with increasing turbulent field contributions in the case of the PS model. The tendency is however generic: With increasing turbulent field strengths and extensions, the sizes of regions of extreme magnification and demagnification shrink and eventually disappear. The amplification distributions tend to peak more around the $\mathcal{A} \sim 1$ value. For all three regular GMF models, the green lines ($B_0=1\,\mu$G, $z_{0}=3$\,kpc) are still close to the red lines (no turbulent field), and the blue lines ($B_0=4\,\mu$G, $z_{0}=3$\,kpc) show a visible decrease of the sizes of the extreme (de-)magnification regions, as well as an increase of bins with more moderate $\mathcal{A}$. In all three panels, blind regions completely disappear for the largest turbulent field contribution ($B_0=10\,\mu$G, $z_{0}=10$\,kpc). All distributions are centered on moderate $\mathcal{A}$ bins, but their widths depend on the regular GMF model. The width for the Sun08 model magenta distribution is larger than for those of the two other models. This is due to the large strength of this field in the halo -up to $10\,\mu$G, which is comparable to that of the turbulent field. Regions with $-2 \leq \log_{10}(\mathcal{A}) \leq -1.5$ and with $0.5 \leq \log_{10}(\mathcal{A}) \leq 1$ indeed disappear when a more reasonable regular field strength in the halo is taken, such as for the Sun08-MH model.

\section{Can all UHECR measured at Earth come from a sole nearby extragalactic source?}
\label{OneSource}

In this section we test the hypothesis that all UHECR detected at Earth may 
be due to one nearby source such as Cen~A. Cen~A has been suggested to be a source of UHECR since a long time~\cite{Cavallo,Romero:1995tn,Anchordoqui:2001nt}.

Since we are sufficiently
far from the most prominent objects of the nearby Large Scale
Structure (LSS) and since Cen~A is only 3.4~Mpc away and located within a
relatively cold region of the local LSS, the extragalactic 
magnetic fields are unlikely to cause large deflections between 
Cen~A and the Earth. We will neglect them in the following. Therefore, we only have to consider deflections in the GMF. We plot in Fig~\ref{CenA_Reg_TF} (left panel) the image of Cen~A emitting iron nuclei with $E \geq 55$\,EeV, deflected in the PS regular GMF model only. The blue dots correspond to the arrival directions of such events, and the black disk denotes Cen~A position on the sky. The image is convolved with the Pierre Auger Observatory exposure. We use the exposure formula of Ref.~\cite{Sommers:2000us} with $\delta=-35.2^{\circ}$ for the Auger declination and $\theta_{\rm m}=60^{\circ}$ for its maximum
zenith angle. We backtrace antinuclei from the Earth to
outside the Galaxy, with initial directions following the exposure. The antinuclei energies follow the energies of the 69 events published by the Pierre Auger Collaboration~\cite{correlation:2010zzj}.
Once the antinuclei leave the Galaxy, we save those with
escape directions on the sky within 5 degrees from the center of
Cen~A. This corresponds to assuming that Cen~A is a $\sim5^{\circ}$
extended source. We could have used more realistic geometries than a
disk, but this would not significantly change our conclusions.

Reference~\cite{correlation:2010zzj} discussed the $\simeq 18^{\circ}$ radius overdensity of events with $E \geq 55$\,EeV around Cen~A, and Ref.~\cite{Semikoz:2010cc} found that when removing this overdense region, the rest of the sky above 55\,EeV is still compatible with isotropy. This region is surrounded with red circles in Fig~\ref{CenA_Reg_TF}. Refs.~\cite{Gorbunov:2007ja,Gorbunov:2008ef} suggested that the overdensity may be the heavy nuclei image of Cen~A. Fig~\ref{CenA_Reg_TF} (left panel) shows that in the PS regular GMF model, the images of Cen~A are deflected beyond the $\simeq 18^{\circ}$ radius region around Cen~A. In all other tested models, heavy nuclei are also shifted by at least a few tens of degrees from Cen~A. However, this result is model-dependent and one cannot rule out the hypothesis of Refs.~\cite{Gorbunov:2007ja,Gorbunov:2008ef} yet. Reference~\cite{Fargion:2008sp} suggested that the overdensity around Cen~A may be explained by light nuclei emitted by Cen~A. The distance to Cen~A is small enough to avoid problems with photo-disintegration. References~\cite{Semikoz:2010cc,Giacinti:2010ep} proposed that this overdensity may be the image of Virgo emitting heavy nuclei, shifted by the regular GMF.

Fig.~\ref{CenA_Reg_TF} (left panel) shows that iron nuclei emitted by Cen~A cannot be sufficiently spread over the whole sky in regular GMF models to explain \textit{all} Auger data above 55\,EeV. Moreover, we have seen in Section~\ref{ConsequencesSourcesearches} that for the
ranges of turbulent component parameters used there, $\sim 60$\,EeV iron nuclei would not be
spread over the whole celestial sphere. Therefore, the only potential way for
the one source hypothesis to be true is that the turbulent GMF is
very strong and extended. We choose as an extreme case $B_0=10\,\mu$G
and $z_{0}=10$\,kpc. We assume the turbulent field has a Kolmogorov
spectrum with $L_{c}=50$\,pc. These values for the strength and the extension of
this field are deliberately larger than the maximum values one can
find in the literature. For example, one of the largest values
proposed for $z_{0}$ is $z_{0}\sim7$\,kpc~\cite{Beck:2008ty}.

Blue dots in Fig.~\ref{CenA_Reg_TF} (right panel) show the arrival directions of iron nuclei with $E\geq55$\,EeV emitted by Cen~A, when this extreme turbulent component is added to the PS model. The Cen~A image is convolved with the Pierre Auger exposure. Since deflections are dominated by the turbulent GMF in this case, using another reasonable model for the regular GMF would not noticeably affect the results presented below. Due to computing time reasons, we used in this section a turbulent field generated via the FFT method, see Section~\ref{MethodsField} for more details. In order to have sufficient statistics for the following study, we backtraced $3.6\times10^6$ particles from the Earth.

The turbulent field starts to be sufficiently strong to spread the arrival directions of Cen~A nuclei over the whole celestial sphere. However, even with such a field, the arrival directions of cosmic rays
still display a non-negligible dipolar pattern on the sky. The density
of cosmic ray arrival directions is larger in the half-sky centered on
Cen~A than in the opposite half-sky. This tends to be in conflict with the angular distribution found from the observational data.

\begin{figure}
  \includegraphics[width=0.46\textwidth]{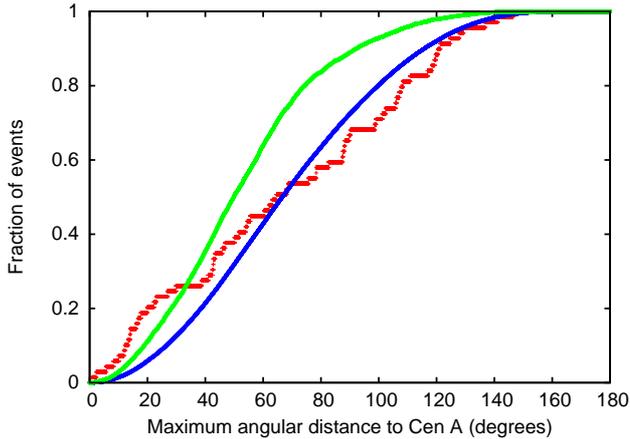}
  \caption{Fractions of cosmic rays within a given
  angular distance (in degrees) to Cen~A. Blue distribution for
  isotropic arrival directions, red one for the 69 Auger events
  published in Ref~\cite{correlation:2010zzj}, and green one for iron nuclei with $E\geq55$\,EeV
  emitted by Cen~A and deflected in the Galactic magnetic field
  model considered for Fig.~\ref{CenA_Reg_TF} (right panel). Both blue and green
  distributions take into account the exposure of the Pierre Auger
  experiment.}
  \label{KStest}
\end{figure}

In Fig.~\ref{KStest} we draw the distributions of
the fractions of cosmic rays within given
angular distances from Cen~A, both for the published Pierre Auger data
with $E\geq55\,$EeV~\cite{correlation:2010zzj} ($N_1=69$ events) and for
$N_2\simeq6300$ cosmic rays emitted by Cen~A and deflected in our extreme GMF
model. These two distributions are respectively represented by the red
and green curves in Figure~\ref{KStest}.
The blue curve on the same panel corresponds to the case of isotropic
arrival directions of cosmic rays (computed with $N_2=10^5$
events). Both blue and green lines take into account the exposure of
the Pierre Auger experiment. The green line is always above the blue
one. This confirms the presence of a non-zero dipolar pattern in the
arrival directions of the cosmic rays emitted by Cen~A.

We verify with a Kolmogorov-Smirnov test~\cite{NR} that the hypothesis
that all cosmic rays are emitted by Cen~A (green distribution) is not
compatible with the Pierre Auger data~\cite{correlation:2010zzj} (red
distribution). This test consists in measuring the maximum distance
$d_{\max}$ between the two distributions. The significance level of
the observed value $d_{\max}$, $P(d_{\max}>\mbox{observed})$, is
directly inferred from $d_{\max}$, $N_1$ and $N_2$~\cite{NR}. The maximum distance between the red and green curves is $d_{\max}\simeq0.28$, at an angular distance to Cen~A $d_{\rm CenA} \simeq 87^{\circ}$. It corresponds to a probability $P(d_{\max}>\mbox{observed})\sim (2-4)\times10^{-5}$.

For a weaker or less extended turbulent GMF than the extreme case considered here, events would be less spread on the sky, and the probability $P(d_{\max}>\mbox{observed})$ would be even lower. Therefore, the hypothesis that Cen~A is the sole UHECR accelerator seen at Earth at the highest energies is strongly disfavoured by the
Pierre Auger data, even if we consider the most favourable case of a heavy composition and a very strong turbulent GMF component.

\section{Conclusions and perspectives}
\label{Conclusions}

In the present paper, we have investigated the consequences of a turbulent
GMF component on the propagation of UHECRs for the case of a heavy
composition at the highest energies. We have backtraced 60\,EeV iron
antinuclei in GMF models including both a regular and a turbulent
component. For the regular component we have mostly used the Prouza and Smida
(PS) model as a generic example representative for regular GMF
models. We have tested the model-dependence of our results by also considering the Sun08 and Sun08-MH models, as in Ref.~\cite{Giacinti:2010dk}. We have varied the parameters describing the turbulent field,
accounting thereby for the poor knowledge of its properties.

In Section~\ref{ConsequencesSourcesearches}, we have computed the
60\,EeV iron images of the Fornax galaxy cluster and of the
supergalactic plane. We have shown qualitatively to which extent the
turbulent component may spread these images. We have also discussed
the dependence of such a spread on the turbulent GMF parameters.

In Section~\ref{MagnificationBR}, we have shown that the presence of a
turbulent field tends to reduce the extreme (de-) magnification of
individual source fluxes. 
We have called ``blind regions'' those parts of the sky in which the
flux of UHECR sources is demagnified by more than a factor
100. Current and next generation experiments will not be able to
detect sources located in such regions. The size of these regions
shrinks when including the effect of a turbulent GMF: At 60\,EeV, for
sufficiently strong turbulent field strengths of $\simeq4\,\mu$G at
Earth and large extensions $\gtrsim3\,$kpc into the halo, the fraction
of blind regions was reduced by more than a factor 3 in the case of
the PS regular GMF model.

In the last section, we tested the hypothesis that all UHECR detected
at Earth above $\simeq55\,$EeV could be due to a sole nearby source,
such as Cen~A. We found that, even in the most favourable case of iron
nuclei deflected in a strong and extended GMF turbulent component,
this scenario is not compatible with the present data of the Pierre
Auger experiment.

Finally we note that future radio experiments, such as LOFAR and SKA,
will improve our knowledge of the properties of the GMF turbulent
component, such as its strength and its extension in the Galactic
halo~\cite{Gaensler:2004gk,Beck:2004gq}.

\section*{Acknowledgments}

This work was supported by the Deutsche Forschungsgemeinschaft through
the collaborative research centre SFB 676. GS and GG acknowledge
support from the State of Hamburg, through the Collaborative Research
program ``Connecting Particles with the Cosmos'' within the framework
of the LandesExzellenzInitiative (LEXI).





\end{document}